\documentclass[preprint,superscriptaddress]{revtex4-1}
\usepackage{amsmath}
\usepackage{graphicx}
\usepackage[dvipdfm]{color}
\begin{document}
\title{An SU(${\cal N}$) Mott insulator of an atomic Fermi gas realized by large-spin Pomeranchuk cooling}
\author{Shintaro Taie}
\altaffiliation{Electronic address: taie@scphys.kyoto-u.ac.jp}
\affiliation{Department of Physics, Graduate School of Science, Kyoto University, Japan 606-8502}
\author{Rekishu Yamazaki}
\affiliation{Department of Physics, Graduate School of Science, Kyoto University, Japan 606-8502}
\affiliation{CREST, JST, 4-1-8 Honcho Kawaguchi, Saitama 332-0012, Japan}
\author{Seiji Sugawa}
\affiliation{Department of Physics, Graduate School of Science, Kyoto University, Japan 606-8502}
\author{Yoshiro Takahashi}
\affiliation{Department of Physics, Graduate School of Science, Kyoto University, Japan 606-8502}
\affiliation{CREST, JST, 4-1-8 Honcho Kawaguchi, Saitama 332-0012, Japan}
\date{\today}
\begin{abstract}
The Hubbard model, containing only the minimum ingredients of nearest neighbor hopping and on-site interaction
for correlated electrons, has succeeded in accounting for diverse phenomena observed in solid-state materials.
One of the interesting extensions is to enlarge its spin symmetry to SU(${\cal N}>2$), which is closely related to systems
with orbital degeneracy.
Here we report a successful formation of the SU(6) symmetric Mott insulator state with an atomic Fermi gas of ytterbium (${}^{173}$Yb)
in a three-dimensional optical lattice. Besides the suppression of compressibility and the existence of charge excitation gap
which characterize a Mott insulating phase,  we reveal an important difference between the cases of SU(6) and SU(2) in the
achievable temperature as the consequence of different entropy carried by an isolated spin. This is analogous to Pomeranchuk
cooling in solid ${}^3$He and will be helpful for investigating exotic quantum phases of SU(${\cal N}$) Hubbard system
at extremely low temperatures.
\end{abstract}
\maketitle

In the last decade, a great deal of progress has been made for two-component atomic Fermi gases.
High controllability and simplicity for these systems allow systematic study over extremely
wide range of system parameters, including interatomic interactions.
One of the milestone experiments in strongly correlated regime are the recently reported
realization of a fermionic Mott insulator \cite{Joerdens2008,Schneider2008} for atoms in optical lattices,
which is of interest itself and also the parent state of high-$T_c$ superconductors \cite{Lee2006}.

On the other hand, many-body physics with multi-component Fermi gases is experimentally unexplored
despite of increasing theoretical interests \cite{Ho1999,Wu2003,Honerkamp2004,Cherng2007,Hermele2009,Cazalilla2009,Gorshkov2010,Yip2011}.
Fermionic isotopes of alkaline-earth-metal like atoms, such as ytterbium (${}^{173}$Yb) \cite{Fukuhara2007a}
and strontium(${}^{87}$Sr) \cite{DeSalvo2010, Tey2010}, are suitable for this aim because of their simple
SU(${\cal N}=2I+1$) symmetric interactions for nuclear spin $I$ \cite{Wu2003,Cazalilla2009,Gorshkov2010}.
${\cal N}$-component Fermi gas with SU(${\cal N}$) symmetry in an optical lattice is well described by
the SU(${\cal N}$) Hubbard model
\begin{eqnarray}
H = -t \sum_{\langle i,j \rangle , \sigma} (c^\dagger_{i,\sigma}c_{j,\sigma} + \text{H.c.})
+ \frac{U}{2} \sum_{i,\sigma \neq \sigma'} n_{i,\sigma} n_{i,\sigma'}
+ \sum_{k=x,y,z} V_k \sum_{i,\sigma} \left(\frac{k_i}{d} \right)^2 n_{i,\sigma}, 
\end{eqnarray}
where $c_{i,\sigma}$ is fermionic annihilation operator for site $i$ and spin $\sigma= -I,\cdots ,+I$,
$n_{i,\sigma}=c^\dagger_{i,\sigma} c_{i,\sigma}$ is the number operator, $V_k = m\omega_k^2 d^2 / 2$
is the strength of harmonic confinement along $k(=x,y,z)-$axis with an atomic mass  $m$ and trap frequency $\omega_k$,
and $d$ denotes the lattice constant.
All parameters, hopping matrix element $t$, on-site interaction $U$ and confinement $V_k$ are
independent of spin states, which manifests the SU(${\cal N}$) symmetry of the system.
Positive scattering lengths $a=10.55$ nm for ${}^{173}$Yb \cite{Kitagawa2008} and $a=5.09$ nm for ${}^{87}$Sr
\cite{Escobar2008} correspond to repulsive interactions ($U>0$), the case of interest in the context of most theoretical studies.
Low temperature behavior of the SU(${\cal N}>2$) model is predicted to be qualitatively different from that of SU(2) model,
mainly due to the enhancement of quantum fluctuation for large ${\cal N}$ system \cite{Wu2006}.
For instance, the analysis on the square lattice, often considered in the Heisenberg limit $U/t \rightarrow \infty$,
reveals the tendency toward disordered spin states \cite{Affleck1988,Read1989,Wu2006,Kawashima2007,Hermele2009,Hermele2011},
whereas the ground state of the half-filled SU(2) model is widely believed to be N\'{e}el-ordered.
Moreover, a striking difference is theoretically predicted for a one-dimensional system.
An infinitely small repulsive interaction results in a formation of a Mott insulating state 
for SU(2) system, whereas a finite strength is required for SU(${\cal N}>2$) \cite{Assaraf1999,Manmana2011}.
Experimental study of SU(${\cal N}>2$) system will lead to better understanding of the underlying physics, and also
will provide insights into the important role of the orbital degeneracy \cite{Li1998,Tokura2000} in condensed matter physics.

A milestone in the study of the SU(${\cal N}>2$) Hubbard system is the realization and characterization of the SU(${\cal N}$)
Mott insulating phase. There are several signatures for a successful formation of a Mott insulating state.
One important feature is the existence of the charge excitation gap. As the interaction $U$ increases, the density of states at the
Fermi level is decreased, which suppresses the mobility of atoms. Finally the system enters the incompressible Mott phase
when the gap opens. At the same time, multiple occupation of lattice sites becomes energetically unfavorable and suppressed
in the Mott regime. More quantitatively, fraction of atoms in doubly occupied lattice sites (double occupancy) is closely related
to the compressibility $\partial n/\partial \mu$ evaluated at the trap center \cite{Scarola2009}. 
Besides these common features of a Mott insulator, it is especially important to clarify the difference between the behaviors
of Mott insulators with SU(2) and SU(${\cal N}>2$) symmetries.

Here, we report a successful formation of the SU(${\cal N}=6$) symmetric Mott insulator state with an six-spin component
atomic Fermi gas of ${}^{173}$Yb in a three-dimensional optical lattice. From double occupancy measurements
with photoassociation spectroscopy and
lattice modulation spectroscopy, we confirm the above characteristics of the Mott state.
Precise control of the spin degrees of freedom provided by optical pumping enables us a straightforward comparison between
the cases of SU(6) and SU(2). We find an important difference that a lower temperature is obtained for SU(6) Mott insulator
as the consequence of larger entropy carried by isolated spin \cite{Cazalilla2009,Hazzard2012} (See also discussions in Section \ref{section_dpr}).
In particular, at the lowest temperature achieved, the entropy density at the center of the trap reaches $\ln (6)$,
which originates from spin degrees of freedom.
Our experimental results are in good agreement with a theoretical calculation based on a high-temperature series expansion (HTSE)
which is reliable in the parameter regime of the current experiment \cite{Scarola2009,DeLeo2011,Hazzard2012} and a local density approximation
(LDA) accompanied with continuum approximation to take into account the presence of the harmonic confinement (See Methods).
This work is an important first step and opens the door to the new frontier of the study of strongly correlated phases
of the SU(${\cal N}>2$) system.

Experimental procedure is as follows (see also Methods for details).
The sample is prepared by loading evaporatively cooled Fermi gas of ${}^{173}$Yb into the optical lattices with simple
cubic geometry. The initial temperature before loading to the lattice is around 20\% of the Fermi temperature $T_{\rm F}$.
In the following sections, we specify the initial condition in terms of corresponding entropy per particle $s$, in the unit
of the Boltzmann constant $k_{\rm B}$. Double occupancy is measured using photoassociation (PA) technique
\cite{Rom2004,Sugawa2011a}. We focus on the Mott phase with unit filling, namely  one atom per lattice site, and
average density at the trap center is below 2 for all experiments presented here.
In this case, we can neglect multiple occupation $n_i \geq 3$ and double occupancy is simply related to the atom loss 
$N_{\rm loss}$ induced by PA, as $D=N_{\rm loss}/N$ where $N$ is the total atom number without PA.
%
%
%
\section{Lattice Modulation Spectroscopy of SU(6) Fermions} 
\begin{figure}[htb]
\includegraphics[width=80mm]{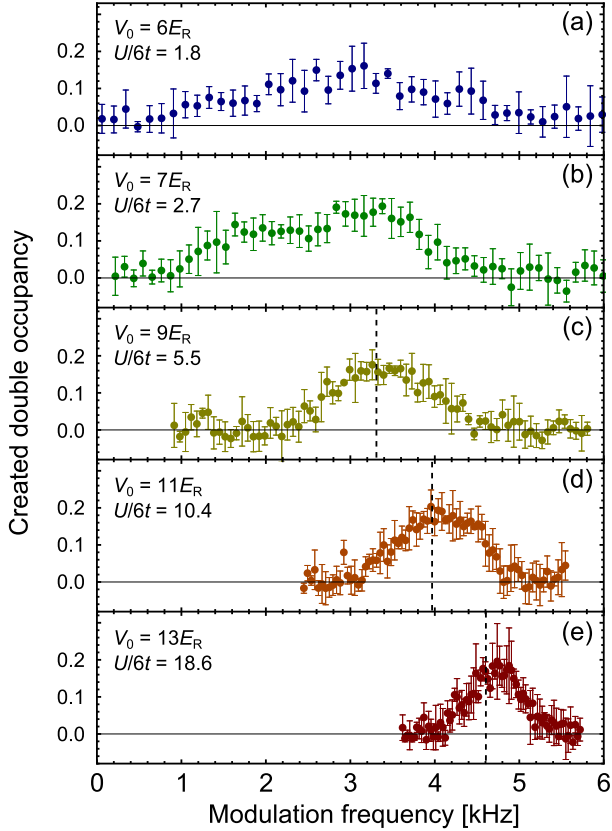}
\caption{
{\bf Lattice modulation spectroscopy.}
Lattice Modulation spectra obtained for samples with $N=1.9(1) \times 10^4$ and $s_{\rm init} = 1.9(2)$,
modulation time of $0.4 h/t$, and amplitudes of $\delta V/V_0 = 0.125,\ 0.115,\ 0.10,\ 0.090,\ 0.085$ for lattice depths of
$6,\ 7,\ 9,\ 11,\ 13$ in unit of $ E_{\rm R}$, respectively.
The vertical dashed lines indicate the calculated values of on-site interaction $U/h$ for the corresponding lattice depth.
The error bars denote s.d. of the measurements.}
\label{graph_modspec}
\end{figure}

First, we present the experimental evidence of the gap of the SU(6) Mott insulator.
The gap can be directly probed by periodically modulating the lattice depth, which induces resonant tunneling
to the occupied lattice sites at the modulation frequency close to the Mott gap $\sim U$ \cite{Joerdens2008}.
This kind of tunneling is detected as the increase in double occupancy.
Figure \ref{graph_modspec} shows the change in double occupancy after lattice modulation, measured at several lattice depths.
Here we apply lattice modulation $V(\tau) = V_0 + \delta V \sin (2 \pi f_{\rm m} \tau)$ with a duration $0.4 h/t$.
Modulation amplitudes $\delta V$ are chosen to set a perturbation strength $F = \delta t/t - \delta U/U$
\cite{Reischl2005} to be $-0.30$. The Hubbard parameters $t$ and $U$ are calculated using the formulae given by
Gerbier et al \cite{Gerbier2005}.
The Mott gaps are clearly observed especially at higher lattice depths. 
For the lattices of $V_0 \geq 9 E_{\rm R}$, we find that the observed spectrum is well fitted by Gaussian and the peak positions
agree with the calculated value of $U/h$ within 3\%.
%

Lattice modulation spectroscopy does not only give information about excitation spectrum but also correlation between
nearest-neighbor lattice sites \cite{Greif2011}. 
From the perturbative treatment for the time evolution of the system under lattice modulation \cite{Greif2011,Kollath2006,Huber2009,Hassler2009},
we obtain the sum rule about the doublon production rate (DPR) $\varGamma (f_{\rm m})= h/t \partial D/\partial \tau$
\begin{equation}
\int \varGamma (f_{\rm m}) {\rm d}f_{\rm m} = 12 \pi^2 F^2 \frac{t}{h} {\cal P},
\end{equation}
where ${\cal P} = N^{-1}\sum_{\langle i,j \rangle} {\cal P}_{ij}$ is the system averaged correlator.
In the case of low filling $n \lesssim 1$ with strong repulsive interaction, the nearest neighbor correlator ${\cal P}_{ij}$ is expressed as
\begin{equation}
{\cal P}_{ij} = \sum_{\sigma \neq \bar{\sigma}}
\left\langle n_{i,\sigma} \prod_{\sigma' \neq \sigma}(1-n_{i,\sigma'}) n_{j,\bar{\sigma}}
\prod_{\bar{\sigma}'\neq \bar{\sigma}} (1-n_{j,\bar{\sigma}'}) \right\rangle,
\end{equation}
where spins $(\sigma, \sigma', \bar{\sigma}, \bar{\sigma}')$ take ${\cal N}$ values.
For small modulation amplitudes, observed DPR shows quadratic dependence on $F$, which justifies
the use of above linear response theory.
In the high-temperature regime $T \gg t^2/U$, we can neglect the spin correlation between adjacent sites
and the above expression for ${\cal P}_{ij}$ reduces to
\begin{equation}
{\cal P}_{ij} = \frac{{\cal N}-1}{{\cal N}} W_i(1) W_j(1),
\end{equation} 
where $W_i(1)$ denotes the probability of single occupation at site $i$. 
The factor $({\cal N}-1)/{\cal N}$ is nothing more than the effect of Pauli exclusion on the hopping process. 
Compared with the case of SU(2), larger DPR is expected for an SU(6) Mott insulator because of factor $5/3$
enhancement due to the reduction of the Pauli exclusion effect. On the other hand, the thermodynamic properties are
reflected in the correlator through $W_i(1)$, which increases with lowering temperature and approaches unity
when the system becomes a defect-free Mott insulator with unit filling.

\begin{figure}[htb]
\includegraphics[width=140mm]{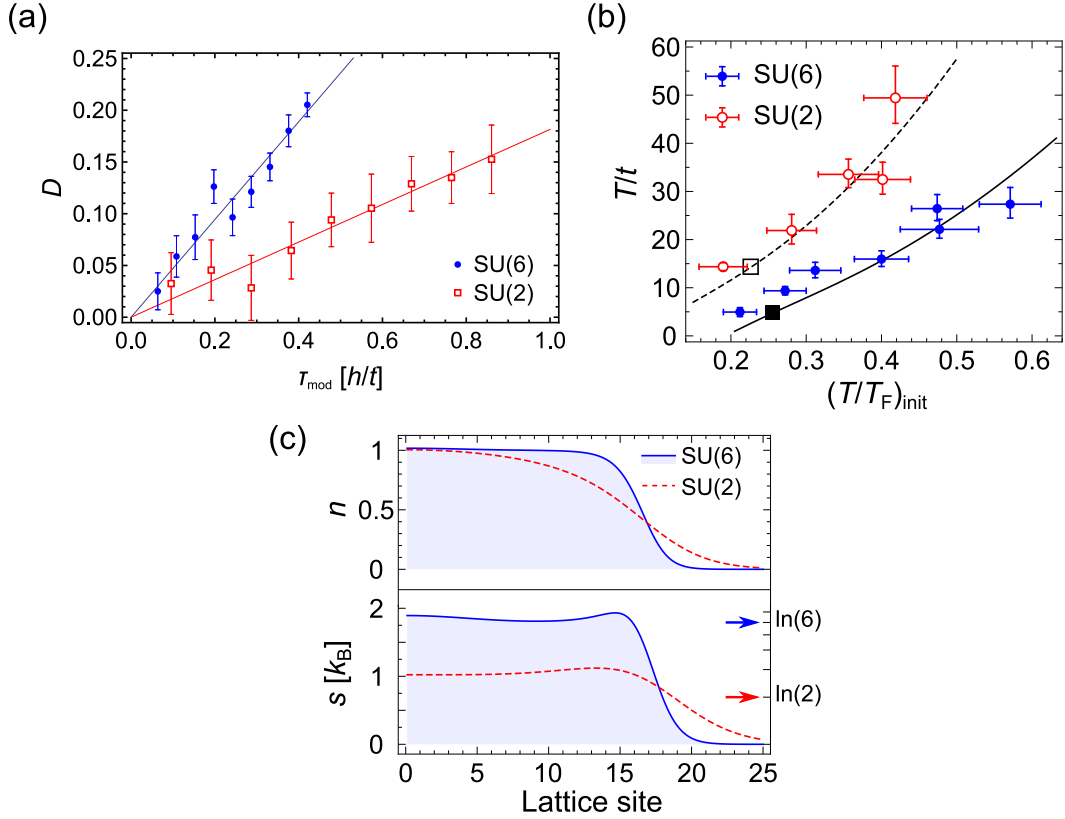}
\caption{
{\bf Doublon production rate for SU(${\cal N}$) Mott insulators.}
(a) Measurement of the doublon production rate. Linear increase is observed for short modulation times presented here.
Atom number is $N=1.9(1) \times 10^4$, the lattice depth is $11E_{\rm R}$ ($t/h=63.7$ Hz and $U/h=4.0$ kHz),
and the modulation amplitude $\delta V/V_0$ is 9.0\%. 
Linear fits yield the values $\varGamma=0.47(2)$ and $0.18(1)$ for the case of SU(6) and SU(2), respectively.
Error bars denote s.e. for 10-15 independent measurements.
(b) Temperatures of SU(6) (blue circles) and SU(2) (open red circles) Fermi gases in the lattice inferred from the measured DPR.
The dependence on the initial temperature in the harmonic trap is shown.
The solid and dashed lines are the corresponding theoretical curves which assume adiabatic loading into the lattice.
Error bars include the fitting error in determining DPR and s.d. of initial temperatures.
(c) Calculated density (top) and entropy distribution (bottom) at the lowest measured temperatures for 6- and
2-component cases, indicated by square symbols in the graph (a). Maximum spin entropy $\ln({\cal N}=6,\ 2)$ are
indicated by the arrows.}
\label{graph_dpr}
\end{figure}

\section{Comparison Between SU(6) and SU(2) Fermions}\label{section_dpr}
We measure the DPR at the peak of the modulation spectrum, both for SU(6) and SU(2) Mott insulators,
as shown in Fig. \ref{graph_dpr} (a) (see also supplementary information S1).
Here we produce SU(2) symmetric system of ${}^{173}\text{Yb}$ by optical pumping \cite{Taie2010} (See Methods), 
in order to make a direct experimental comparison between the behaviors of SU(6) and SU(2) systems.
The modulation spectra are well fitted with a Gaussian shape of $e^{-2}$ full width of $25(2) t/h$, determined from
the spectrum for modulation time $\tau=0.3h/t$ (see Fig. S1 (a)).
The peak DPR is extracted from linear fitting to the data taken for several modulation times.
The measured DPR for an SU(6) Mott insulator exceeds $5/3$ of that for SU(2),
which indicates the lower temperature is realized (Fig. \ref{graph_dpr} (a)).
This is clarified in Fig. \ref{graph_dpr} (b), where we plot the temperature in the lattice determined from the measured DPR.
Temperature estimation is based on the formalism presented in the previous section and the calculation of the quantity $W_i(1)$
from the high-temperature series expansion (HTSE), as described in Methods.
The temperatures for SU(6) Fermi gases are the factor of about $2 \mathchar`-3$ smaller than that for the SU(2) cases.
We also show in Fig. \ref{graph_dpr} (c) the density and entropy distributions calculated for the case of the largest DPR, in other words,
the lowest temperature achieved.
In the case of SU(6), a robust Mott plateau is clearly formed and entropy per site is very close to $\ln (6)$.
We note that, in a trapped system, entropy is pushed onto a metallic state near the edge of the cloud and
a Mott insulator at the trap center survives for higher total entropy \cite{Schneider2008}. 
On the contrary, starting from the almost same initial $T/T_{\rm F}$ in a harmonic trap before loading into the lattice,
the plateau has been largely melted for the SU(2) case.

%
\begin{figure}[htb]
\includegraphics[width=100mm]{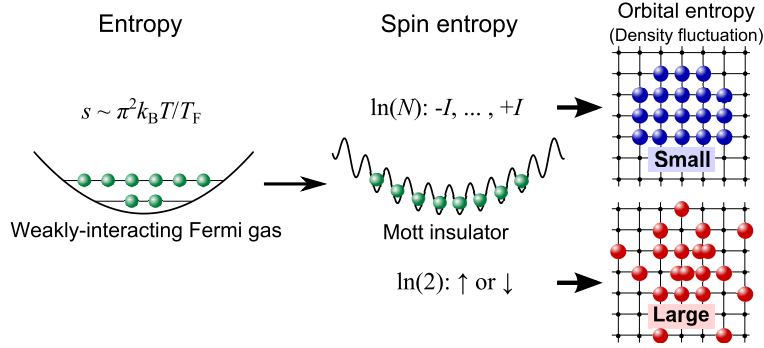}
\caption{
{\bf Schematic of enhanced Pomeranchuk cooling in an SU(${\cal N}$) Fermi gas.}
When the system evolves from a weakly-interacting degenerate Fermi gas into a Mott insulator,
spin degrees of freedom arise.
Lager entropy can be absorbed by isolated spins of an SU(${\cal N}>2$) Mott insulator, resulting in
the reduction of density fluctuation which has dominant contribution in determining the temperature.
In real experiments, the behavior of absolute temperature strongly depends on the harmonic confinement:
the system can even be heated if strong compression occurs during loading into the lattice.
However, the general trend that larger ${\cal N}$ leads to lower temperature remains unchanged.
}
\label{graph_schematic}
\end{figure}

The observed striking difference between the temperatures required for accessing the Mott phase of SU(2) and SU(${\cal N}>2$)
cases can be understood by a following simple argument, as shown in Fig. \ref{graph_schematic}.
In the atomic limit where tunneling between lattice sites is neglected, the maximum spin entropy per atom is
equal to $\ln {\cal N}$ for an ${\cal N}$-component Fermi gas in the Mott state with unit filling.
On the other hand, motional degrees of freedom must be frozen in the Mott phase. 
Assuming adiabaticity of loading process into optical lattices, we can expect that the defect-free Mott phase in an optical lattice
forms at $(T/T_{\rm F})_{\rm init} \sim \ln {\cal N} / \pi^2$ where $(T/T_{\rm F})_{\rm init}$ is
measured in a harmonic trap before loading into lattices.
This implies a significant reduction of the required initial temperature in the unit of $T_{\rm F}$ for large ${\cal N}$ system,
as predicted in Refs. \cite{Cazalilla2009,Hazzard2012}. 
In other words, large spin can effectively cool down the system by absorbing entropy from motional degrees of freedom,
which is the same mechanism as Pomeranchuk cooling observed in solid ${}^3$He \cite{Richardson1997}.

Although this mechanism was responsible for achieving novel quantum phases in a mixed gas of bosons and fermions
in a recent report \cite{Sugawa2011a}, no systematic study was done on the detailed behaviors of enhanced Pomeranchuk cooling,
especially the comparison between the SU(2) and SU(6) which is clearly demonstrated in this work.

\section{Double Occupancy Measurement of SU(6) fermions } 
%
\begin{figure}[htb]
\includegraphics[width=140mm]{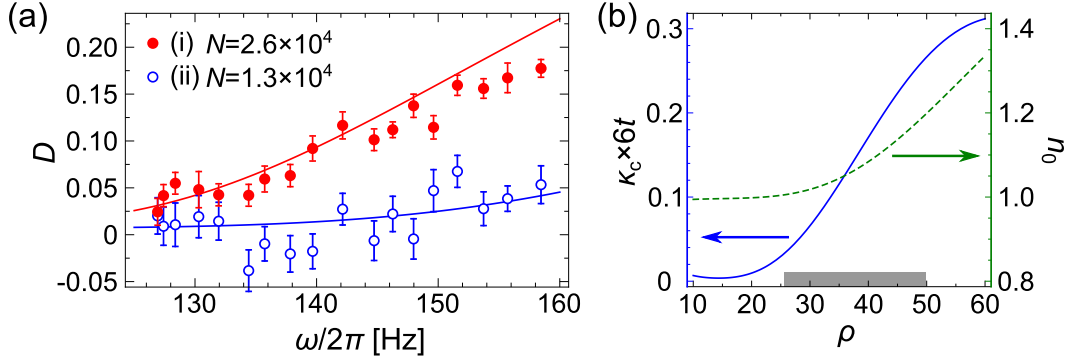}
\caption{
{\bf Double occupancy and compressibility.}
(a) Measured double occupancy. Two data series with different atom number
(i) $N=2.6(1) \times 10^4$ (red circles) and (ii) $1.3(1) \times 10^4$ (open blue circles) are shown.
The solid lines are the best fitted theoretical curves according to the second order high-temperature
expansion, corresponding $s=2.2$ and $2.5$ for (i) and (ii), respectively.
The error bars denote s.e. for typically 15 independent measurements.
(b) Calculated central compressibility $\kappa_{\rm c}$ (solid) as the function of characteristic filling $\rho$,
together with the central density $n_{\rm c}$ (dashed) with the parameters for data (ii) in (a).
Fixed entropy $s=2.5$ obtained by the fit in (a) is used. Experimentally relevant range of $\rho$ is indicated by the shaded region.
}
\label{graph_do}
\end{figure}

Double occupancy measurement has been extensively used as the probe for fermionic lattice
systems \cite{Joerdens2008,Schneider2008,Joerdens2010} to obtain the information on the compressibility.
While the correlation measurement presented above is a good probe for the low density case $n \leq 1$, the double occupancy
is suitable for characterizing states with high filling, especially $1 \leq n \leq 2$.
Figure \ref{graph_do} (a) shows the measured double occupancy for various mean trap frequencies
$\omega = (\omega_x \omega_y \omega_z)^{1/3}$, at the lattice depth of 9.0$E_{\rm R}$. 
Corresponding Hubbard parameters are given as $t/h=101$ Hz and $U/h=3.3$ kHz.
The data are taken for two initial preparations of  (i) $N=2.6(1)\times 10^4$ and $s_{\rm init}= 1.7(1)$ and
(ii) $N=1.3(1)\times 10^4$ and $s_{\rm init}= 1.9(3)$.
For the data (i), the double occupancy increases with tighter confinement and the trap center is metallic with $n>1$.
On the other hand, the data (ii) shows that the double occupancy is essentially zero for low compression regime,
indicating the formation of an $n=1$ Mott plateau.
We fit the data to a theoretical curve based on the HTSE by taking the entropy per particle in the lattice $s$ as only
one fitting parameter. We then obtain $s=2.2$ and $2.5$ for the set (i) and (ii), respectively, which are slightly higher
than $s_{\rm init}$. Since they still lie between $s_{\rm init}$ and the entropies measured after reversed loading into
the harmonic trap which range from $2.8$ to $3.4$, the deduced values of entropy $s$ are reasonable and consistent
with the existence of some constant heating or non-adiabatic effects during the loading and reverse loading processes. 

Incompressibility due to the repulsive interaction, which characterizes a Mott insulator, can be deduced from the
obtained entropy in the lattice. In Fig. \ref{graph_do} (b) we show the calculated density $n_0$ and compressibility
at the trap center $\kappa_{\rm c}=\partial n_0 / \partial \mu$ as a function of characteristic filling defined as
$\rho = N(12t/m \omega^2 d^2)^{-3/2}$.
Introducing characteristic filling enables to unify the data with different total atom number and the harmonic confinement,
within the LDA formalism \cite{Rigol2004,DeLeo2011}.
For the lowest measured $\rho \sim 25$, $n_0$ is almost unity and $\kappa_{\rm c}$ is as low as $0.03 / 6t$,
which means that the system is deep in the strongly incompressible Mott insulating regime.
We note that this value of $\kappa_0$ is a factor of $\sim 3$ smaller than that expected in the case
of the SU(2) Hubbard model with the same parameters ($s$, $\rho$, and $U/t$), as the consequence of
large-spin Pomeranchuk cooling mentioned above.
%

This work clearly shows the realization of an SU(6) Mott insulator with unit filling and demonstrates
the existence of enhanced Pomeranchuk cooling for large spin systems.
Pomeranchuk cooling is most effective for systems with low filling for which atoms possesses full spin degrees of freedom.
We note that it also works for higher fillings because (a) entropy per atom
$n^{-1} \ln ({}_{\cal N}{\rm C}_{n})$ is larger than $\ln(2)$ for $n \leq {\cal N}/2$ and
(b) a Mott plateau with high filling is surrounded by shells with lower filling, where spin degree of freedom survives.

\subsection*{{\it Note added.}}
After completion of our work, we noticed some theoretical works \cite{Cai2012,Messio2012,Bonnes2012} which study
thermodynamics of SU(${\cal N}$) Hubbard/Heisenberg model. They found that the advantage of large-${\cal N}$
systems in adiabatic cooling, which is demonstrated in our experiment, is still valid in low temperature regime
($k_{\rm B}T \sim t^2/U$) where magnetic correlations take place.
\section{Methods}    
\subsection{Preparation of a degenerate fermi gas of ${}^{173}$Yb with 6 spin components}
The method for preparing a degenerate Fermi gas of ${}^{173}{\rm Yb}$ is essentially the same as
that described before \cite{Fukuhara2007a} except that the detailed parameters of the experimental apparatus
have been changed (For current setup, see also Ref. \cite{Sugawa2011b}).
A balanced mixture of all nuclear spin states is evaporatively cooled in a crossed far-off resonant optical trap (FORT).
At the end of evaporation, we have $1-3 \times 10^{4}$ atoms with a temperature around $0.2\, T_{\rm F}$.
Temperature is obtained by performing standard Thomas-Fermi fitting to the observed momentum distributions
and entropy per particle is calculated from measured $T/T_{\rm F}$ using the formula for non-interacting Fermi gas.
The spin distribution is measured by an optical Stern-Gerlach effect and confirmed to be equal within 5\% relative uncertainty
\cite{Taie2010}.
Atoms are subsequently loaded into an optical lattice within 150 ms. The lattice potential is formed by three mutually
orthogonal laser beams with the lattice constant $d=266$ nm. During the first 100 ms of loading process,
the lattice is ramped up to $5E_{\rm R}$ and at the same time we change the power of the FORT laser to
obtain desired confinement strength.
The lattice depth is calibrated by analyzing diffraction patterns of a Bose-Einstein condensate of
${}^{174}{\rm Yb}$ after the application of the pulsed lattice potential \cite{Denschlag2002}.
Trap frequencies are determined from dipole oscillations. For lattice modulation experiments (Fig. \ref{graph_modspec}
and \ref{graph_dpr}), the mean trap frequencies $\omega/2\pi$ are $115.2(8),\ 119.6(9),\ 127.4(10),\ 134.5(11),\ 140.9(13)$ [Hz]
for the lattice depths of $6,\ 7,\ 9,\ 11,\ 13$ in unit of $E_{\rm R}$, respectively.
\subsection{Preparation of a degenerate Fermi gas of ${}^{173}$Yb with 2 spin components}
At the early stage of evaporative cooling, atoms are pumped into
$m_F = +5/2$ and $-5/2$ states by $\pi-$polarized laser light which is resonant with the
${}^1S_0 \leftrightarrow {}^3P_1 (F'=3/2)$ transition. The fraction of residual spin components are estimated
to be $N_m \leq 0.03 N_{+5/2}$ for $m=-3/2,\cdots +3/2$. We observe no detectable spin changing collision such as
$m_F = (+5/2, -5/2) \rightarrow (+3/2, -3/2)$ during several second of evaporation, which is the characteristic
of SU(${\cal N}$) symmetry \cite{Gorshkov2010}.
\subsection{Measurement of doubly occupied site by photoassociation}
First, the lattices are ramped up to desired depth for which Mott insulating states can be reached.
We then increase the lattice depth to $25E_{\rm R}$ in 2 ms to suppress tunneling during the measurement,
followed by 10 ms irradiation of PA laser light. The PA laser is detuned by -796 MHz from the ${}^1S_0 \leftrightarrow {}^3P_1$
atomic transition ($\lambda = 556 \text{\ nm}$) and has the intensity of $\sim 0.5  \text{W/cm}^2$. 
The PA process enables to convert all atoms on doubly occupied sites into electronically excited molecules which rapidly escape
from the trap. The loss of atom is, therefore, the measure of the double occupancy.
\subsection{Theoretical method}
Theoretical calculations presented in this paper are based on high-temperature expansion of the
SU(6) Hubbard model \cite{tenHaaf1992,Henderson1992,Hazzard2012}. 
Extending the calculation in Ref. \cite{Henderson1992} to the SU(${\cal N}$) case,
we have the grand-canonical free energy per lattice site up to second order in $t/k_{\rm B}T = \beta t$:
\begin{eqnarray}
\Omega (T, \mu) &=& \Omega_0 (T, \mu) + \Delta \Omega, \\
- \beta \Delta \Omega &=& \left( \frac{\beta t}{Z_0} \right)^2 z {\cal N} \left[ \frac{1}{2} \sum _{n_1=1}^{\cal N}
\binom{{\cal N}-1}{n_1 - 1}^2 \right. x^{2n_1-1} w^{(n_1-1)^2}  \nonumber \\
&&\left. -\frac{1}{\beta U} \sum_{n_1 \neq n_2}^{\cal N} \binom{{\cal N}-1}{n_1-1} \binom{{\cal N}-1}{n_2-1}
\frac{x^{n_1+n_2-1}w^{\frac{1}{2} n_1(n_1-1)+\frac{1}{2} (n_2-1)(n_2-2)}}{n_1-n_2} \right],
\end{eqnarray}
where $z$ is the number of nearest neighbors, $x={\rm e}^{\beta \mu}$,
$w={\rm e}^{-\beta U}$, and $\Omega_0$ is the free energy in the atomic limit given by
\begin{eqnarray}
\Omega_0 (T, \mu) &=& -k_{\rm B}T \ln Z_0 (T, \mu), \\
Z_0 (T, \mu) &=& \sum_{n=0}^{\cal N} \binom{\cal N}{n} \exp \left[-\beta \left( \frac{U}{2}n(n-1) - \mu n\right) \right].
\end{eqnarray}
For a trapped system, we apply local density approximation in which the system is regarded as locally uniform.
Thus we have $\Omega_{\rm LDA} = \sum_i \Omega_i= \sum_i \Omega(T, \mu_i)$, where
$\mu_i = \mu - V_i$ is the local chemical potential for the external potential $V_i$.
Finally we approximate the sum over lattice sites with spatial integration (continuum approximation).

As our experiments are carried out near unit filling $n=1$ with strong repulsive interaction, we can neglect
multiple occupation with filling $n \geq 3$. In this case, double occupancy is given by the same formula as
for the SU(2) model:
\begin{equation}
D = \frac{2}{N} \sum_i \langle n_{i,\uparrow} n_{i,\downarrow} \rangle
= \frac{2}{N} \frac{\partial \Omega_{\rm LDA}}{\partial U}.
\end{equation}
To obtain the expression for the correlator ${\cal P}_{ij}$, we use the relation
$\langle n_i \rangle = \sum_n^6 n W_i(n) \simeq W_i(1) + 2W_i(2) = -\partial \Omega_i / \partial \mu$, where
$W_i(2)$ is local double occupancy and is given by $\partial \Omega_i / \partial U$.
Then the correlator is given in terms of the derivatives of the free energy as
\begin{equation}
{\cal P}_{ij} = \frac{{\cal N}-1}{{\cal N}} \left( \frac{\partial \Omega_i}{\partial \mu}+2\frac{\partial \Omega_i}{\partial U}\right) 
\left( \frac{\partial \Omega_j}{\partial \mu}+2\frac{\partial \Omega_j}{\partial U}\right).
\end{equation}

\subsection*{Acknowledgements}
We acknowledge M. A. Cazalilla, T. Giamarchi, V. Gurarie, K. R. A. Hazzard, M. Hermele, K. Inaba, A. M. Rey, A. Tokuno and M. Yamashita for helpful discussions.
This work is supported by the Grant-in-Aid for Scientific Research of JSPS (No. 18204035, 21102005C01 (Quantum Cybernetics)), 
GCOE Program "The Next Generation of Physics, Spun from Universality and Emergence" from MEXT of Japan,
and World- Leading Innovative R\&D on Science and Technology (FIRST).
ST acknowledge supports from JSPS.


%

%

%

%

\clearpage
\setcounter{figure}{0}
\begin{center}
{\large \textbf{Supplementary Information}}
\end{center}
\renewcommand{\figurename}{FIG. S}
\newcommand{\figureref}[1]{Figure S\ref{#1}}
\section*{S1. Measurement of nearest neighbor correlator}
\begin{figure}[hb]
\includegraphics[width=140mm]{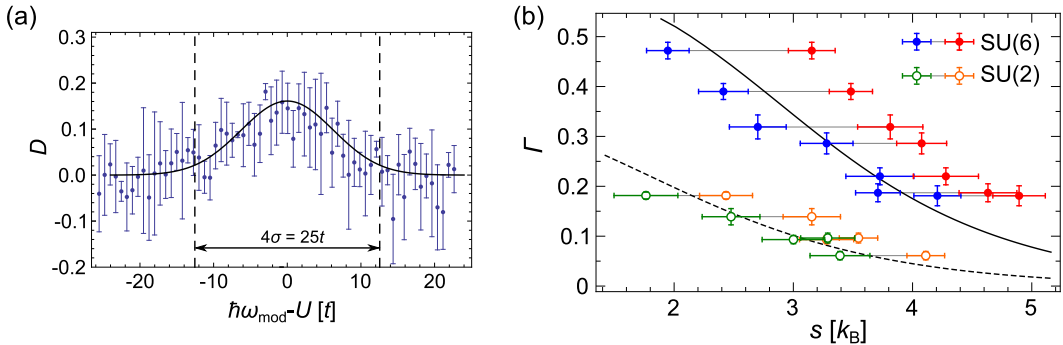}
\caption{
(a) Lattice modulation spectrum for the lattice depth of $11E_{\rm R}$,
the modulation amplitude of $\delta V/V_0=9.0\%$, and modulation time of $0.32 h/t$,
which is shorter than that for Fig. 2. We extract the width $4\sigma=25(2) t$
from a Gaussian fit to the data. Error bars denotes s.d. of the measurements.
(b) Peak DPR obtained for both SU(6) (closed circles) and SU(2) (open circles) Mott insulators,
as a function of entropy per particle $s$ measured in the harmonic trap.
We measure $s$ before loading to the lattice (blue and green) and also after reversing the loading prededure (red and orange),
which set the lower and the upper limit, respectively. The corresponding two values are connected by the gray lines.
}
\label{graph_rate}
\end{figure}

Within the perturbative formalism used in this paper, we need the frequency-integrated
doublon production rate (DPR) to obtain the nearest neighbor correlator. Reference \cite{Greif2011spl} reported that
$\varGamma (\hbar \omega) $ has approximately Gaussian shape with ${\rm e}^{-2}$ full width of $24t$ for a trapped
SU(2) Mott insulator.
In our case of an SU(6) Mott insulator, the spectrum has the width of $25(2) t$ (\figureref{graph_rate} (a)),
which is in good agreement with their result.
Given the width of a modulation spectrum, the correlator can be calculated from the DPR measured at
the peak of the spectrum.
\figureref{graph_rate} (b) shows the peak DPR measured at several initial temperatures,
which is the basis of thermometry presented in Fig. 2 (b) in the main text. Remarkable enhancement of
the DPR for SU(6) fermions is observed, which is due to the combined effect of the greater Pauli suppression of tunneling
for the SU(2) case and the enhancement of the Pomeranchuk effect for the SU(6) case.

\clearpage
\section*{S2. Effect of residual spin components}
Here we discuss the effect of residual population in the $m_F = \pm 3/2$ and $\pm 1/2$ states due to imperfect optical pumping
to $m_F = \pm 5/2$ states in the case of SU(2) experiments.
For a uniform Mott insulator with unit filling, the entropy per atom is given by
$s = k_{\rm B} N^{-1}\ln (N!/ \prod_\sigma N_\sigma!)$ where $N$ is the total atom number and $N_\sigma$ is the
atom number for the spin component $\sigma$.
Below we consider the special case that each minority component is equally populated.
We define the imbalance parameter $p$ as $N_{-3/2} = \cdots = N_{+3/2} = p N_{+5/2} (= p N_{-5/2})$.
In the case of $p=0.03$ which is upper limit in our experiment, we have $s/k_{\rm B} = 0.96$, considerably
larger than $\ln(2)=0.69$ for the pure SU(2) case. This large spin entropy cools the sample and enhances
DPR. In addition, impurity components reduce the effect of Pauli principle in hopping processes, which also
contributes to the increase of DPR. We estimate this effect to be $\sim 10\%$ for $p=0.03$.

\begin{figure}[htb]
\includegraphics[width=85mm]{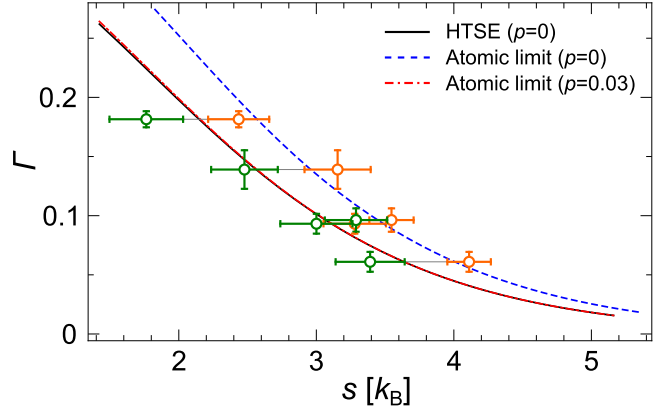}
\caption{
Effect of the impurity spin components on the doublon production rate.
The experimental data (green and orange circles) and the HTSE calculation with $p=0$ (solid line) are
identical with that shown as the SU(2) case in \figureref{graph_rate} (b).
The blue dashed line shows the atomic limit calculation with $p=0.03$ described in this section.
Note that, the calculations on the HTSE and the atomic limit for $p=0$ (dash-dotted line) give almost the same result
for the deep lattice ($V_0=11E_{\rm R}$) considered here.
}
\label{graph_imbalance}
\end{figure}

For more quantitative argument, we calculate the DPR for six-component gases with imbalanced population within the atomic limit.
Imbalanced population leads to two chemical potentials $\mu_{\rm m}$ and $\mu_{\rm r}$ for
two majority components of $m_F = \pm 5/2$ and residual minority ones of $m_F = \pm 3/2, \pm 1/2$, respectively.
We consider the probability $W_i(n_{\rm m},n_{\rm r})$ that site $i$ is occupied by $n_{\rm m}$ majority spins and
$n_{\rm r}$ minority spins. In the atomic limit, it is given by
\begin{eqnarray}
W_i(n_{\rm m},n_{\rm r}) &=& \frac{1}{Z} \binom{2}{n_{\rm m}}\binom{4}{n_{\rm r}} B_i (n_{\rm m},n_{\rm r}), \\
B_i(n_{\rm m},n_{\rm r}) &=& {\rm e}^{-\beta \left[ \frac{U}{2}(n_{\rm m}+n_{\rm r})(n_{\rm m}+n_{\rm r} - 1) - \mu_{i,{\rm m}}n_{\rm m} -\mu_{i, {\rm r}} n_{\rm r}\right]},
\end{eqnarray}
where $Z = \sum_{n_{\rm m}=0}^2 \sum_{n_{\rm r}=0}^4 \binom{2}{n_{\rm m}}\binom{4}{n_{\rm r}} B_i(n_{\rm m},n_{\rm r})$
is the partition function with local chemical potentials $\mu_{i,{\rm m}/{\rm r}} = \mu_{{\rm m}/{\rm r}}-V_i$.
Recalling that the correlator is the probability that the site $i$ and $j$ are singly occupied
except for the case of the same spin component, we have
\begin{equation}
{\cal P}_{ij} = \frac{1}{2} W_{i}(1,0) W_{j}(1,0) + \frac{3}{4} W_{i}(0,1) W_{j}(0,1)
+ W_i(1,0) W_j(0,1)+ W_i(0,1) W_j(1,0),
\end{equation}
where the first term represents the case that both sites $i$ and $j$ are occupied by majority
components (${\cal N}=2$ in Eq. (3)), the second term for the case of minority components (${\cal N}=4$),
and the other two terms for the cases that a site is occupied by a majority atom and the other site by a minority one.
\figureref{graph_imbalance} shows the calculated DPR for the imbalanced gas, together with the experimental
results and the HTSE calculation. Although the increase in the calculated DPR due to the impurity components is
clearly visible, it is still within the range of uncertainty of the experimental data.

\end{document}